# Afforestation in Romania: Realities and Perspectives

Ciprian Palaghianu

Stefan cel Mare University of Suceava, Forestry Faculty, Romania
Corresponding author e-mail address: cpalaghianu@usv.ro (C. Palaghianu)

**Abstract:** Romania's forest cover is at this time below, but quite close to the European average. Despite that, recent forestry activities tend to be more oriented towards forest exploitation and not to increase the national forested area. The general public perception is that afforestation activities are limited only to NGO's projects and media actions and numerous reports and statements made by some of these organizations are rather sensational and confusing. This paper tries to cast more light on this controversial issue, presenting accurate data and considering the whole context of afforestation - from the history of afforestation activities in Romania to the recent statements and reports submitted by the government agencies involved in these particular issues. The necessity and opportunity of afforestation is also substantiated from the perspective of reintroduction into production of marginal and degraded land. An analysis of the situation of funding through the National Rural Development Programme (2007-2013) measures is performed and possible future directions regarding afforestation programmes are discussed.

## 1. Introduction

Indisputable, the forest represents a vital global resource. Being a renewable resource it is continuously harvested for quite a considerably amount of time. It is unthinkable to imagine a world without forest because of the many implications of it in our everyday lives. Forest does not embody only a sum of trees, but a complex natural system highly interconnected with a lot of different environmental elements.

Today, we all know the whole benefit package that comes along with the existence of forest in our life. Forest can deliver a wide range of ecological services regarding biodiversity, soil protection, water quality, habitat providing, along with social and recreational services. The new environmental realities recognize the significant role of forests in mitigating climate change, as natural carbon sinks and as a source of renewable wood that can be used as fuel or raw material for different products. Considering these particular aspects, it is easy to understand why forest cover represents such an important indicator in many recent statistics and reports.

The distribution of forest resources is uneven, both at the level of states and continents. Today, the global forest cover is approximately 4 billion hectares, which means that almost one third of the terrestrial area is forested (FAO, 2010). But nearly 8000 years ago, the forest cover was double, according to World Resources Institute data. The development of agriculture and the construction of the modern human society lead to massively loss of forest cover. This phenomenon was concentrated excessively in the last two centuries and the recent trends are uncomforting (Palaghianu, 2009).

The global forest cover has been drastically decreasing annually by nearly 7 million hectares in the last two decades (see table 1). Today we are facing a rapid pace of deforestation and this tendency is likely to get worse considering the increasing trend of wood or wood products consumption (Palaghianu, 2007).

However, in recent years there is a growing interest for wood resources management and remarkable progress became visible considering the afforestation efforts. It's worth mentioning the positive example of Europe which extended its forested area with nearly 1 million hectares per year in the last two decades.

The importance of afforestation in balancing the forest resources is now generally accepted and every regional or national forest strategy includes an afforestation programme (Mather, 1993). The European Union is actively involved in the management of forest resources using the Common Agricultural Policy (CAP) and since 1992 EU is also engaged in afforestation initiatives (Council Regulation 2080/92). Beginning with the year 2000, the forestry sector was incorporated into the Rural Development Plan, according to the Council Regulation 1257/1999.

The UE aims to extend the forested areas and the funding mechanisms support two categories of forest initiatives: afforestation and other forestry measures.

## 2. A review of afforestation activities in Romania

Romania has consistent forest resources: about 6.5 million hectares and the forest cover is estimated at 27% of the country's total area (INS, 2013). However, Romania's forest cover is still below the European Union average which is currently estimated at 42% (EC, 2013).

It is known that Romanian forest cover was superior far back in the past and several well-known studies indicate that fact (Giurescu, 1975; Doniță et al., 1992).

Table 1. Forest cover of the world (according to FAO data, 2010)

| Year | 1990 | 2000 | 2005 | 2010 |
|---|---|---|---|---|
| | million hectares ||||
| Africa | 749 | 709 | 691 | 674 |
| Asia | 576 | 570 | 584 | 593 |
| Europe | 989 | 998 | 1.001 | 1.005 |
| North and Central America | 708 | 705 | 705 | 705 |
| Oceania | 199 | 198 | 197 | 191 |
| South America | 946 | 904 | 882 | 864 |
| **Global** | 4.168 | 4.085 | 4.061 | 4.033 |

During the last two centuries, when the forest loss at global level was at the highest level, Romania lost nearly 2 million hectares of forest, but in the last century the forest cover remained almost unchanged, varying around 6.5 million hectares (see fig. 1).

Taking into account the historical and current loss of the forests, afforestation and reforestation represented a key element in the effort of preserving the forest resources. And such efforts were made from early time in Romania.

We could start with the first document of Grigore Ureche, the chronicler who mentioned the early "*afforestation*" actions guided by Stephen the Great, voivode of Moldavia, after the battle of the Cosmin Forest (1497). More consistent information regarding not only the basic forest management but also afforestation was specified in the forest regulations from Transylvania (1775 and later in 1781). These first guidelines and procedures were quickly followed by similar forest protocols in Bukovina (1786), Moldavia and Wallachia (1792).

However, the earliest afforestation actions should be considered the mobile sand fixation and afforestation that were executed in Oltenia province in 1852 (Giurescu, 1975). Soon after that, in 1864 the first three forest nurseries were established, each of it having an area of 50 hectares (in Brăila, Iași and Ismail counties).

The promulgation of the first Forest Code in 1881 brought some consistent improvements in the forestry sector, and later, in 1889 a new national service was founded, a service that joined afforestation and torrents control.

This action demonstrates the early understanding of the role of afforestation in controlling torrents, landslides and erosion.

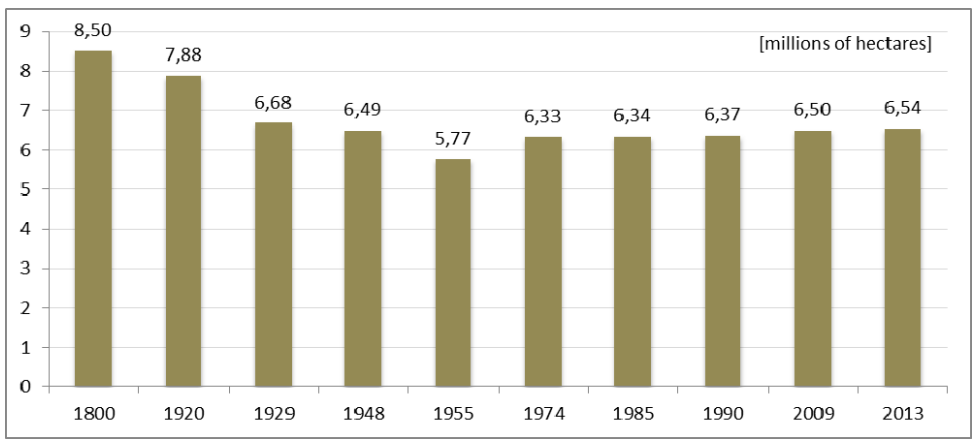

Figure 1. Changes of the forest cover in Romania during the last two centuries (unit: millions of hectares)

Specific tasks were undertaken in the effort of preventing and controlling the erosion of watershed in the following years and later at the beginning of the 20th century.

For instance, in a rather difficult period, between 1930 and 1947, 97.000 hectares of degraded land were afforested (Crăciunescu et al, 2014).

After the World War II, the Romanian forests were nationalized and the state took over the forest resource management. The WWII reparations paid to Soviet Union and the intense activity of the SOVROM, Soviet-Romanian enterprises, created to manage the debt recovery, have left deep scars in Romanian forests. In the 50's the forest condition was severely altered - more than 700 thousand hectares were deforested. Furthermore, an additional 600 thousand hectares were considered degraded lands. As a result, a massive afforestation national plan was developed for reforestation of 1 million hectares. This notable objective was finally achieved in 1963, after more than a decade of impressive afforestation efforts. The annual average of reforested land was around 70 thousand hectares, with a maximum of 98 400 hectares in 1953.

In the communist period, the forestry management was quite well balanced and manifested a particular interest in afforestation. In order to direct and control the field specialists there were edited Technical Guidelines in 1948, 1953, 1966, 1969, 1977 and 1987. The standards for seed quality assessment were updated in 1963, 1973, 1983 and the afforestation infrastructure was developed by creating numerous seed orchards, seed stands, seed processing facilities and large forest nurseries in order to boost the seedlings production.

## 3. Present state of afforestation in Romania and future trends

Today, the Romanian forest covers 6.5 million hectares (INS, 2013) but after 1990 the state was no longer the sole landowner and manager. Still, the state represents, with nearly 50% of the forest, the most important landowner and the legal state-owned entity, RNP - National Forestry Administration, represents the largest administrator of Romanian forest.

After 1990 the forestry sector suffered remarkable changes. The planning and control of the state were not so strict due to the change of the regime. Unfortunately, new obstacles appeared, considering the forest fragmentation and restitution. The infrastructure needed for afforestation was also damaged because forest nurseries and seed orchards were returned to former owners, who have neglected or destroyed it.

The new forestry paradigm was better focused on natural regeneration and the afforestation effort was abruptly reduced. Moreover, illegal logging and forest fragmentation have contributed to the negative general public perception on forestry.

Unfortunately common people consider many of the significant silvicultural activities to be unfamiliar and incomprehensible, because of an improper dissemination. Furthermore, quite few public statements clarify the available data and information, which are generally ignored by the public (Palaghianu & Nichiforel, 2016).

Although significant changes in forest cover were not found (Dutcă & Abrudan, 2010; MECC, 2010; INS, 2013), it is quite obvious that Romanian forests have suffered at least structural alterations in the past two decades.

This observation is emphasised by numerous media campaigns – some of them extremely persuasive and partially controversial. Greenpeace, WWF Romania or the local initiative "*Plantăm fapte bune în România*" were heavily involved in such environmental campaigns that pointed out to the recent loss of forests. It's worth mentioning the Russian Greenpeace report on Romanian forest (Greenpeace, 2012), which stated that 3 hectares of forest per hour are "*disappearing*". Numerous media entities have cited this report and its results and many associated this loss of forest with illegal logging, which in fact was not accurate. The public perception was easily altered by media pressure and by misunderstanding the difference between authorised clear-cutting needed in forest regeneration and illegal logging. In association with the increased wood logging and undersized afforestation plans (see figure 2) the general perception on Romanian forestry seems that it is more oriented to logging, whatever these actions are legit or not.

However, what is the reality, beyond the media slogans or persuaded perceptions? Although, the TBFRA-2000 report (TBFRA, 2000) presented a positive average annual change of forest of 14.7 thousand hectares between 1955 and 1990, this was not a constant trend. After 1990 the rhythm of afforestation was clearly diminished. The reports on the first decade after 1990 (Georgescu & Daia, 2002), indicated an annual average of nearly 10 thousand hectares of afforestation between 1992 and 2001 and a comparable value for the natural regenerated areas. This trend regarding afforestation was maintained at the almost same level for the next period (2005-2013), as shown in figure 2, considering only the afforestation completed by RNP.

After 1990, the state was not the only forest administrator and different funding mechanisms emerged. Despite all the property and administration changes, RNP still represents the most active and visible entity involved in the afforestation effort. RNP has several different mechanisms of funding afforestation: its own budget, the Fund for the improvement of the lands with forest destination, the Fund for forest conservation and regeneration, as well as funds from the state budget. The output of all funding instruments leads to the previously mentioned annual average of nearly 10 thousand hectares afforested.

In the past two decades there were additional funding mechanisms that have been available or used for afforestation programs: the SAPARD (Special Accession Program for Agriculture and Rural Development) Program – measure 3.5 (for the period 2000-2006), the Environmental Fund and Environment Fund Administration afforestation programs and, finally, the European Union funding instruments that have been implemented by the National Program for Rural Development (PNDR) measures (in 2007-2013 and 2014-2020 programs).

The SAPARD Program (2000-2006) was implemented by PNADR (National Plan for Agriculture and Rural Development) and the measure 3.5 was specifically designed for forestry.

The 3.5 measure included more than 7 million euros (7.445 mil. euros) funding support for afforestation (a target of 15,000 hectares) and nearly 4 million euros (3.722 mil. euros) for forest nurseries. At the end of the program, 3 afforestation projects and one nursery project were funded and a disappointing 1.3% funding absorption rate was reached (MADR, 2011).

Unfortunately, the Environmental Fund (founded by government emergency ordinance no.196/2005) and Environment Fund Administration were not able to produce more significant results. Due to excessive bureaucracy and numerous changes in the funding guides, one single afforestation project (designed for an area of 40.5 hectares) was funded in the first 7 years from the creation of this funding mechanism (RCA, 2013).

The latest updated reports show an improvement in the past years: 2,836 hectares of afforestation were funded till 2014 (RCA 2014).

Considering the partial failure of the previous funding mechanisms, the European Union funding instruments were considered more adequate and better balanced. The National Program for Rural Development (PNDR) for 2007-2013, granted 1.2 billion euros for forestry measures of the total 7.5 billion euros.

The 221 Measure – *The first afforestation of agricultural land* was designed for promoting afforestation projects, with a total budget of 229 million euros.

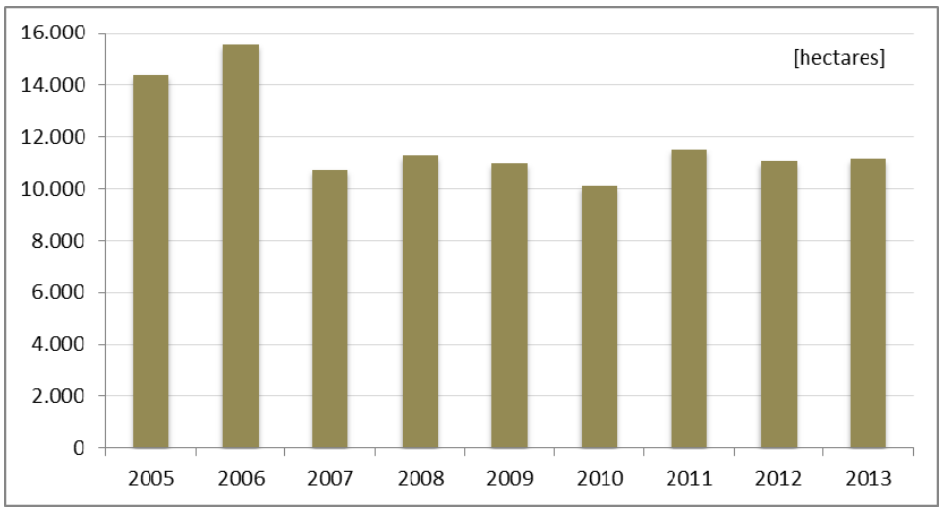

Fig. 2 Afforestation completed by the National Forestry Administration RNP (2005-2013) (unit: hectares)

The outcomes at the end of the term consisted in 29 projects, funded with a total of 185 thousand euros, resulting in a shocking absorption rate of 0.08%.

Another initiative was the Measure 122 - *Improving the economic value of forests*, which encompasses various activities from the forestry field, including the creation of nurseries. From the total of 135 million euros, only 1 million euros was used in funding projects and consequently the absorption rate was 0.8%.

The new PNDR 2014-2020, granted only 300 million euros from the total 8 billion euros (excluding the 10 billion euros for direct payments) for forestry measures. The Measure 8.1 for afforestation has a budget of 105 million euros and doubles the standard costs for the afforestation activities, in order to boost the absorption rate of funds for this particular field.

## 4. Conclusion

It is quite obvious that investments in afforestation do not seem very attractive due to the considerable time gap between investments and benefits, in this particular field.

However, considering the important role of forests in the environmental and social paradigm as well in the mitigation of climate change, the state should play a more significant role. In the absence of private investors, the state should be more active and should encourage afforestation by different funding mechanisms or attractive tax incentives.

The UE funding scheme is ineffective without an adequate support from the state. We can mention the Mediterranean states (Spain, Italy or Portugal) which have greatly benefited from the UE funds (Zanchi et al, 2007). Their outcomes regarding afforestation were as solid as their afforestation policies (Palaghianu & Clinovschi, 2007).

In Romania, the official statements recognize the importance of afforestation and one important objective of the National Afforestation Programme (2004) and the Forest Code (2008) was the afforestation of 2 million hectares of degraded lands. Moreover, the Law no. 100 /2010 regarding the afforestation of degraded lands was a new reinforcement of that ambitious objective, but the new versions of the National Afforestation Programme from 2010 and 2013 altered successively the target from 2 million hectares to 422 thousand hectares, respectively to 229 thousand hectares.

Although, the latest results of Romania in the field of afforestation seem inconclusive, different opportunities will soon arise along with the development of the new PNDR 2014-2020.

The latest Measure 8.1 for afforestation appears to be limited in budget, but it brings new mechanisms of payment and the promise of reducing the bureaucracy. This might be a fresh restart for the old Romanian afforestation engine.

# References

Crăciunescu, A., Moatăr, M., & Stanciu, S. (2014). Considerations regarding the afforestation fields. Journal of Horticulture, Forestry and Biotechnology, 18(1), 108-111.

Doniţă N, Ivan D, Coldea G, Sanda V, Popescu A, Chifu T, Paucă-Comeănescu M, Mititelu D, Boșcaiu N. (1992). Vegetaţia României. D. Ivan (Ed.). Ed. Tehnică Agricolă, 407 p.;

Dutcă, I., & Abrudan, I. V. (2010). Estimation of Forest Land Cover Change in Romania between 1990 and 2006. Bulletin of the Transilvania University of Brasov, Series II-Forestry, Wood Industry, and Agricultural Food Engineering, 52(1), 33-36.

Food and Agriculture Organization (FAO) of the United Nations. (2010). Global forest resources assessment 2010: Main report. Food and Agriculture Organization of the United Nations.

Giurescu, C.C., 1975: Istoria pădurii românești din cele mai vechi timpuri până astăzi [History of the Romanian forests from ancient times to today], Ed. Ceres, Bucuresti, 388 p.;

Georgescu, F., Daia, M.L., (2002). Forest regeneration in Romania, Proceedings of the seminar"Afforestation in the context of SFM", Ennis, 15 – 19 September 2002, 112-121;

Greenpeace, (2012). Evoluția suprafețelor forestiere din România în perioada 2000 – 2011,(www.greenpeace.org/romania) 5p.;

INS (National Institute of Statistics), (2013). Romanian Annual Statistical Report for the year 2013.

Mather, A. (1993). Afforestation policies, planning and progress. Belhaven Press. 223 p.;

MADR, (2011). Romanian Ministry of Agriculture and Rural Development – Final report on the implementation of SAPARD Programme in Romania, 304 p.;

Palaghianu, C. (2007). Aspecte privitoare la dinamica resurselor forestiere mondiale. Analele Universitatii Stefan cel Mare Suceava - Sectiunea Silvicultura, 9 (2), 21-32;

Palaghianu, C., Clinovschi, F. (2007). Analiza situatiei si tendintelor in ceea ce priveste impaduririle pe mari zone fizico-geografice. Analele Universitatii Stefan cel Mare Suceava - Sectiunea Silvicultura, 9 (2), 33-38;

Palaghianu, C. (2009). Researches on forests regeneration by informatical tools. Universitatea" Ștefan cel Mare", Suceava, PhD Diss.[In Romanian].

Palaghianu, C., Nichiforel, L. (2016). Between perceptions and precepts in the dialogue on Romanian forests. Bucovina Forestiera, 16 (1), 3-8

RCA (Romanian Court of Auditors), (2013). Sinteza Raportului de audit privind situația patrimonială a fondului forestier din România, în perioada 1990-2012. Romanian Court of Auditors Report;

RCA (Romanian Court of Auditors), (2014). Sinteza Raportului de audit al performanței modului de administrare a fondului forestier național în perioada 2010 – 2013. Romanian Court of Auditors Report;

Temperate and Boreal Forest Resource Assessment (TBFRA-2000). (2000). Forest Resources of Europe, CIS, North America, Australia, Japan and New Zealand (industrialized Temperate/boreal Countries): UN-ECE/FAO Contribution to the Global Forest Resources Assessment 2000 (No. 17). United Nations Publications;

Zanchi, G., Thiel, D., Green, T., Lindner, M., (2007). Forest Area Change and Afforestation in Europe: Critical Analysis of Available Data and the Relevance for International Environmental Policies, European Forest Institute, Technical Report 24, pp. 45;